# Effect of substrate temperature on the spin transport property in $C_{60}$-based spin valve devices

Feng Li*

*High Magnetic Field Laboratory, Chinese Academy of Science, Hefei 230031, China*

*lfeng99@mail.ustc.edu.cn

**We report the effect of the substrate temperature on the magnetoresistance (MR) of the $C_{60}$-based spin valve (SV) devices with the sandwich configuration of $La_{0.67}Sr_{0.33}MnO_3$ (LSMO)/$C_{60}$/cobalt (Co). The $C_{60}$ interlayer deposited at different substrate temperatures resulted in four types of devices. We observed all types of devices showed a monotonic increase in MR ratio with the substrate temperature. And more interesting, an especially large MR (28.5%) were obtained in the device fabricated at higher substrate temperature, while for the other types of devices, the MR magnitudes were just about a few percent. Based on the I-V measurements, as well as SEM and AFM characteristics, we have obtained that the higher substrate temperature can cause many pits and hollows in the organic film, and these pits will increase the tunneling probability of spin-polarized carriers from one ferromagnetic electrode to the other.**

**Introduction**

Recently organic semiconductors (OSCs) become a promising candidate as the nonmagnetic spacer in the spin-valve (SV) devices, motivated by their long spin-relaxation lifetimes due to weak spin-orbit interaction and hyperfine interaction [1, 2]. Giant magnetoresistance (GMR) effect has been demonstrated on the vertical SV devices employing the organic spacers such as tris(8-hydroxyquinoline) aluminum ($Alq_3$) and rubrene thicker than 100 nm [3–7]. In parallel, OSC ultrathin layers performed successfully in spin tunnel junctions [8-10], and extremely high (>300%) tunneling magentoresistance (TMR) has been observed at low temperatures [9]. However, in the some reported cases where a clear GMR effect was reported, the organic SV devices with relatively thick OSC spacer display low device resistance and a weak temperature dependence of I-V behavior, characteristic of TMR. This fact raised a serious debate on whether the observed MR in the OSV is due to carrier transport within the organic layer or tunneling through the locally thin areas of the OSC layer [11-13]. It was found that interfacial reaction and metal penetration into the 'soft' organic layer takes place upon the deposition of the ferromagnetic electrode, leading to the formation of an ill-defined layer and pinhole channels within the OSC spacer [3, 14, and 15]. Furthermore, in the most recent experimental works on the organic SV devices [3-20], the organic interlayer mostly had the amorphous structure.

Fullerene $C_{60}$ was chosen as an organic spacer in the vertical SV devices in our work. It is motivated by its very weak hyperfine interaction due to the absence of polarized hydrogen nuclei [16], as well as the good match of lowest unoccupied molecular orbital (LUMO) of $C_{60}$ with the Fermi energy of common ferromagnetic electrodes, which makes a relatively easy carriers injection from the electrodes. Besides, the $C_{60}$ molecules are very robust, unlike other organic materials, can sustain without being damaged upon electrode deposition [17-20].

Here, by changing the substrate temperature during the deposition of $C_{60}$ interlayer, we have improved the crystallinity of $C_{60}$ interlayer and have fabricated the different LSMO/$C_{60}$/Co devices which exhibit the diverse MR valves with the same $C_{60}$ thickness. Furthermore, we have obtained clear evidences spin transport behavior in these devices adopting Current-voltage (I-V), Scanning electron microscopy (SEM), and Atomic force microscopy (AFM) measurements.

**1 Experimental**

The device structure is shown schematically in Figure 1(a). The 100 nm thick and about 1 mm wide LSMO films, which is the half-metallic ferromagnet with nearly 100% spin polarization and also very stable, were deposited on $SrTiO_3$ (STO) <100> substrates using pulsed laser deposition method through a shadow mask. After cleaning using acetone, the substrate with the narrow LSMO film was introduced into the evaporation chamber (based pressure $\leq 1\times10^{-7}$ Torr), where a $C_{60}$ layer was thermally evaporated. The 80 nm thick $C_{60}$ film was vacuum-evaporated with the same evaporation rate, confirmed by the thickness profilometer, but the different substrate temperatures (respectively, 25 ℃, 50 ℃, 100 ℃ and 150 ℃). The growth rate and total thicknesses of $C_{60}$ interlayer in the four sets of devices was set to the same value and *in situ* monitored by a quartz crystal thickness monitor. Finally, The 10 nm thick Co film was thermally evaporated as top ferromagnetic electrode and subsequently capped with a 30-nm Al protection layer using shadow mask. The cross-bar junction area was about $0.5\times1$ mm². all fabrication steps involving the organic layer and the top electrode were performed inside a glove-box or inside the glove-box-integrated vacuum evaporation.

The crystalline phase of the $C_{60}$ films was identified by X-ray diffraction (XRD) using the Rigaku-TTR3 X-ray diffractometer with Cu Kα ($\lambda$= 1.5406 Å) radiation. The morphology of the samples was determined by field-emission scanning electron microscopy (FESEM) (FEI, Model Sirion-200). The morphology of the $C_{60}$ layer was characterized with a Veeco MultiMode (nanoscope V) AFM in tapping mode. The MR values calculated using the expression MR% = $(R_{AP} - R_P) \times 100/R_P$, where $R_{AP}$ and $R_P$ is the resistance of the anti-parallel and parallel magnetization configuration of two electrodes, of the devices were measured in a Quantum Design Physical Property Measurement System (PPMS) under an external in-plane magnetic field from room temperature (300 K) to 4.2 K using the standard four-probe method. Figure 1(a) also shows the experimental setup for a four-probe measurement. Current-voltage (I-V) measurements were performed using a Keithley 2612 source measure unit, with the positive pole connected to the LSMO electrode. Furthermore, the energetic location of the highest occupied molecular orbital (HOMO) and lowest unoccupied molecular orbital (LUMO) of $C_{60}$, together with the Fermi levels ($E_F$) and the work functions of LSMO and Co, are shown in Figure 1(b), which indicate the good match of LUMO of $C_{60}$ with $E_F$ of common ferromagnetic electrodes, which makes a relatively easy carriers injection from the electrodes. The inset shows the chemical structure of $C_{60}$.

**2 Results and discussion**

Figure 2(a), inset, shows a typical magnetoresistance trace for a LSMO/$C_{60}$ (80 nm)/Co spin valve device, the $C_{60}$ interlayer was deposited at the substrate temperature of 25 ℃, measured at 50 K. The figure shows that sizeable magnetoresistance can be achieved in these devices. The main panel of this figure shows that the magnitude of the magnetoresisitance signal decreases with increasing temperature. Furthermore, we can see that all the devices show typical negative MR loop, which is in accordance with the previous report in the organic spin valve devices employing the LSMO and Co as the anode [3-5, 19]. The temperature dependence is believed to reflect a loss in spin injection efficiency from the LSMO [19]. The bias-voltage dependence of the magnetoresistance ratio, for the device prepared at the substrate temperature of 25 ℃, at 50 K is presented in figure 2(b). The asymmetric magnetoresistance ratio for the bias voltage is due to different electrodes (LSMO, Co), and which is similar to the previous reports [3-5, 11, 13].

Figure 3(a) shows the typical MR curves for the organic SV devices, measured at 50 K, with the 80 nm thick $C_{60}$ interlayers prepared at the substrate temperatures of 25 ℃, 50 ℃, 100 ℃, and 150 ℃, respectively. The resistance of the devices decreases monotonously as the substrate temperature. It is about more than 100 k$\Omega$ for device fabricated at the substrate temperature of 25 ℃, while for the device, which was fabricated at the substrate temperature of 150 ℃, exhibits just about several k$\Omega$. It would be very interesting that the organic spin valve devices with the same thick organic interlayer, but just deposited at different substrate temperature, exhibit so much large different resistances. Furthermore, since the fabricated $C_{60}$-based SV devices show the junction resistance larger than 1000 $\Omega$, two orders of magnitude higher than that of LMSO and Co electrode, The MR signal should not be associated with anisotropic magnetoresistance effect in the LMSO and Co electrode.

The MR ratios, calculated based on the resistances of the devices which were prepared at four substrate temperatures in figure 3(a), versus the magnetic field (H) at 50 K are shown in figure 3(b), which are 8.6%, 11.2%, 13.3%, and 27.6% for the devices fabricated at the substrate temperature of 25 ℃, 50 ℃, 100 ℃, and 150 ℃, respectively. It can be shown that the MR ratios increase moderately with increasing substrate temperature at the range of 25 ℃ between 100 ℃, but for the devices prepared at substrate temperature of 150 ℃, the MR ratio increase intensity. In order to further investigate the relationship between the MR ratio and the substrate temperature, we fabricated the $C_{60}$-based SV devices, with the same thick organic interlayer of 80 nm but at different substrate temperatures, and measured the MR ratios of them at 50 K. Figure 3(c) shows that the magnitude of the MR signal increases with increasing substrate temperature. It can be seen clearly further that the MR ratio of the device, prepared at substrate temperature of 150 ℃, is extraordinary large than that of the devices which were fabricated at other substrate temperatures. It is a very fascinating physical phenomenon. We deduce tentatively it may origin from high crystallinity for the $C_{60}$ interlayer fabricated at high substrate temperature, especially at 150 ℃. That is because the high crastallinity is propitious to the transport of the spin polarized carriers, so the spin relaxation length is very long in the high crystalline materials.

In order to further clarify this issue that the reason for the large MR ratio in the device prepared at high substrate temperature, we have performed the specular X-ray diffraction measurements to find out the crystal structure and investigate the structural change with the substrate temperature. The specular XRD patterns of the $C_{60}$ films, deposited on STO/LSMO at four different substrate temperatures to mimic their growth as in the actual spin valve devices, are shown in Figure 4. It can be seen that the X-ray diffraction (XRD) pattern of a $C_{60}$ thin film displays strong peaks that correspond to the (111) and (311) reflections of the face-centered cubic (fcc) phase of $C_{60}$, revealing the crystalline nature of the sample. And every corresponding diffraction peak of the films, prepared at different substrate temperatures, appear at the similar degree, which means the crystal structure of the four films is similar. However, we can find that the corresponding diffraction peaks are very weak for the $C_{60}$ film fabricated at substrate temperature of 25 ℃. It is also to be noted that the diffraction peaks become stronger gradually as the substrate temperature increases, and the diffraction peaks for the film prepared at substrate temperature of 150 ℃ are particularly stronger than those of other films prepared at the lower substrate temperatures. This can verify preliminarily the large MR ratio measured in the device prepared at high substrate temperature could arise from high crystallinity for the $C_{60}$ interlayer deposited at high substrate temperature, and crystallinity generally favors an enhancement in mobility.

In order to get deeper insight on the mechanism of spin transport in $C_{60}$-based spin valves with the 80 nm $C_{60}$ interlayer of different crystalline degree, current–voltage (I-V) characteristics of the SV devices with the same $C_{60}$ thicknesses but deposited at different substrate temperatures were measured at different temperatures and were studied as shown in Figure 5. For the device grown at substrate temperature of 150 ℃, the behavior of the I-V curves are shown in Figure 5(a). It is shown that I-V curves are weakly non-liner and are almost temperature independent, which indicates the evident characteristic of spin polarized carrier tunneling.

The spin polarized carriers transport could be usually described as a sum of two distinct pathways: one is tunneling through the HOMO–LUMO gap and the other is electron injection and then hopping transport within the LUMO levels. The elastic tunneling through the barrier is generally considered to be limited to 2-3 nm of the insulator [21]. At small thickness of the organic semiconductor spacer, the multi-step tunneling current through the defect states inside the gap will serves as a main channel of device current due to the existence of injection barrier at the Co/$C_{60}$ interface. Beyond the tunneling limit, electrons have to be elevated thermally or via the electric field to the LUMO level [11], where electron can hop more easily among the LUMO levels of adjacent molecules. Thus the current will be controlled by carrier injection and transport within the $C_{60}$ layer.

For the $C_{60}$ film we prepared, its thickness is about 80 nm, and spin-polarized carrier tunneling should be unlikely to occur in such a thick organic interlayer. Furthermore, it is worthy to note from figure 5(a) that the device resistances, measured at four different temperatures, for the device prepared at the substrate temperature of 150 ℃ are magnitude of several thousand Ohms, which are about two orders of magnitude lower than those for the device prepared at the substrate temperature of 25 ℃. The behavior of the I-V characteristics for the device prepared at the substrate temperature of 25 ℃ are shown in Figure 5(b), which are however strikingly different. It is shown that the I-V curves have now become largely temperature dependent and strongly nonlinear at low temperatures. In light of the criteria for distinguishing tunneling and spin injection [11-12], these kind of I-V curves, together with the observation of MR feature is obvious indicative of spin polarized injection. So it should be an unintelligible puzzle why the mechanism of the spin polarized transport is so different in the devices with the same thickness of $C_{60}$ interlayer. In the light of the above experimental results and the corresponding analysis, we can deduce that the significantly larger spin dependent transport length in organic interlayer with high crystal would not be regarded as the dominate factor to achieve the exceptionally larger value of MR ratio in the device which was prepared at the substrate temperature of 150 ℃.

Generally, it should be pointed out that highly crystalline structures in the film will form many large crystals at the surface of the film, and which will increase the surface roughness of the film. While for the film with the amorphous-like structure and low crystallinity, it will display relatively homogeneous surface morphology. Large crystals on the surface of the film will potentially lead to pinholes and pits in the vertical devices. The presence of pinholes and pits in the vertical device will increase the possibility of pinhole channels and magnetic metallic impurities within the organic interlayer, as a result of the penetration of metallic atoms into the pits of the organic film during magnetic metallic electrode deposition, and thereby decrease the effective thickness of the organic spacer and make the effective organic interlayer thickness be in the range of spin polarized tunneling. The presence of spin polarized tunneling channels in the $C_{60}$ spacer should contribute more significantly to the occurrence of the extremely large value of MR ratio in our organic SV devices. Such issue merits further investigation.

Based on the above analysis, and with a view to more insight into the origin for the large value of MR ratio, we have performed firstly the SEM measurements to find out the surface morphologies of the 80 nm thick $C_{60}$ films, which were deposited on the STO/LSMO to mimic their growth as in our actual organic SV devices, as shown in figure 6. From figure 6(a), the morphology of the film deposited at the substrate temperature of 25 ℃ is the very smooth; while for the film which was deposited at the substrate temperature of 25 ℃ (see figure 6(d)), the surface morphology is very rough.

Furthermore, we then have performed the AFM measurements to find out the surface morphology of the $C_{60}$ films with the thickness of 80 nm. Figure 7(a) shows the typical surface topographic image of the 80 nm $C_{60}$ layer grown on the LSMO at the substrate temperature of 25

℃ with the scanning size $2\mu m \times 2\mu m$, and its section image is shown in figure 7(b). The film displays a homogeneous surface, and the surface morphology of the film is very smooth. The root mean squared (RMS) roughness of 0.68±0.002 nm. The RMS roughness is independent of film thicknesses (80 nm) and consistent with a molecular size of $C_{60}$ (around 1 nm). Due to the low roughness and high smoothness, the possible presence of the pinholes and filament conduction channels within the $C_{60}$ layers could be excluded. From the section image, it is also to be noted that the maximum roughness is 4.32 nm, which is especially small compared with the thickness of the film of 80 nm.

The typical surface morphology and its section information of the $C_{60}$ film prepared at the substrate temperature of 150 ℃ are shown in figure 7(c), the scanning size is also $2\mu m \times 2\mu m$. It can be seen that the surface of the film is not smoothness but is uniform and continuous with many biggish nanocrystallines. In contrast, the faceted crystalline $C_{60}$ film deposited at the substrate temperature of 150 ℃ exhibits a very rough surface morphology. The average grain size increases upon increasing the annealing temperature. The AFM images reveal that high temperatures support the crystallization of the $C_{60}$ molecules. Figure 4 presents XRD patterns of the $C_{60}$ thin films deposited at the substrate temperature of 25–150 ℃. The $C_{60}$ films subjected to the lower substrate temperature possess amorphous features, whereas that deposited at the substrate temperature of 150 ℃ exhibit the strong diffraction peaks of crystalline $C_{60}$. However, the $C_{60}$ molecule is the typical spherical molecule, so when the $C_{60}$ film was prepared at higher substrate temperature, the grain size would become larger accordingly, and the pronounced grain-boundary effects become apparent, thereby resulting in many tiny protuberances (light areas) and some pits (dark areas) on the surface of the film (see figure 7(f)). The height of protuberances and the deepness of pinholes can be seen from the section images, the maximum values are about 25 nm and 35 nm, respectively. The RMS roughness of the surface is 8.28 nm. The more intuitionistic 3D AFM images ($2\mu m \times 2\mu m$) of the $C_{60}$ films, deposited at the substrate temperature of 25 ℃ and 150 ℃, are shown in figure 7(e) and figure 7(f) respectively, and which provide more details of the plane areas. Obviously, from figure 7(f), the surface is composed of relatively large uniform nanocrystallines, and the pits and hollows with their vertical deepness of about more than 60 nm spread over the scanning size of the film. These pits and hollows on the surface of the $C_{60}$ films may lead to the formation of ill-defined layer in organic spin valves. It is not difficult to understand if the effective organic semiconductor interlayer is not thick enough, the top Co impurities, which penetrate into and fill in these pits and hollows during magnetic metallic electrode deposition, will be very close and will even contact with the LSMO film, and so that the organic spin valves will form the spin polarized tunneling channels and will be even short-cut. As the result, the especially large MR data obtained in the $C_{60}$-based SV device, in which the $C_{60}$ interlayer was deposited at the substrate temperature of 150 ℃, should originate from the tunneling behavior of spin-polarized carriers through the $C_{60}$ spacer with many pits and hollows on the surface of it. Further more works will be performed on controlling crystallinity of the $C_{60}$ films and thus carrier mobility using some other effective methods, to elucidate the relationship between carrier transport and the MR values and spin diffusion length of the $C_{60}$-SVs.

## 3 Conclusions

In summary, we have prepared the vertical LSMO/$C_{60}$/Co SV devices with the same thick $C_{60}$ spacers which were deposited at different substrate temperatures. The interesting larger value of MR ratio was observed in the device of $C_{60}$ spacer prepared at the higher substrate temperature. Detailed MR curves and I-V characterizations of the devices indicate a transition of spin transport behavior from spin-injection and hopping transport inside the $C_{60}$ spacer, which were deposited at low substrate temperatures, into tunneling through the $C_{60}$ layer prepared at high substrate temperatures. The SEM and AFM measurements further illustrate the occurrence of pits and hollows on the surface of $C_{60}$ spacer, deposited at the higher substrate temperatures, will cause the formation of spin tunneling channels in the vertical $C_{60}$–based SV devices.

**Figure captions**

**Figure 1.** (a) Schematic diagram of our spin valve device structure with an SrTiO$_3$ substrate/La$_{0.67}$Sr$_{0.33}$MnO$_3$/C$_{60}$/Co/Al; (b) Schematic band diagram of the organic device shows the Fermi levels (E$_F$), the work functions of the two ferromagnetic electrodes (La$_{0.67}$Sr$_{0.33}$MnO$_3$ and Co), the highest occupied molecular orbital (HOMO) and the lowest unoccupied molecular orbital (LUMO) levels of C$_{60}$.

**Figure 2.** (a) Temperature dependence of the magnetoresistance measured in the organic spin valve with 80 nm thick C$_{60}$ interlayer, the inset shows the magnetoresistance at 50 K; (b) Bias voltage dependence of the magnetoresistance for the organic spin valve measured at 50 K.

**Figure 3.** (a) Magnetoresistance curves of our spin valve devices, which were prepared at different substrate temperatures, at 50 K; (b) Magnetoresistance ratios for the devices in (a); (c) Substrate temperature dependence of magnetoresistance ratio.

**Figure 4.** X-ray diffraction patterns for the C$_{60}$ films, grown on LSMO, at different substrate temperatures.

**Figure 5.** I–V curves at different temperatures for the LSMO/C$_{60}$/Co spin valves with the 80 nm C$_{60}$ layer which were fabricated at the substrate temperature of (a) 150 ℃ and (b) 25 ℃.

**Figure 6.** SEM images of C$_{60}$ films which were fabricated at the substrate temperature of (a) 25 ℃, (b) 50 ℃, (c) 100 ℃, and (d) 150 ℃. The scale bar in each figures is equal to 200 nm.

**Figure 7.** Atomic force microscopy (AFM) images of the C$_{60}$ films deposited on LSMO at the substrate temperature of (a) 25 ℃ and (c) 150 ℃; The profile along the lines highlighted in (a) and (c) are shown in (b) and (d), respectively; (e), (f) The 3D AFM image of C$_{60}$ film in (a) and (c), respectively.

**Figures**

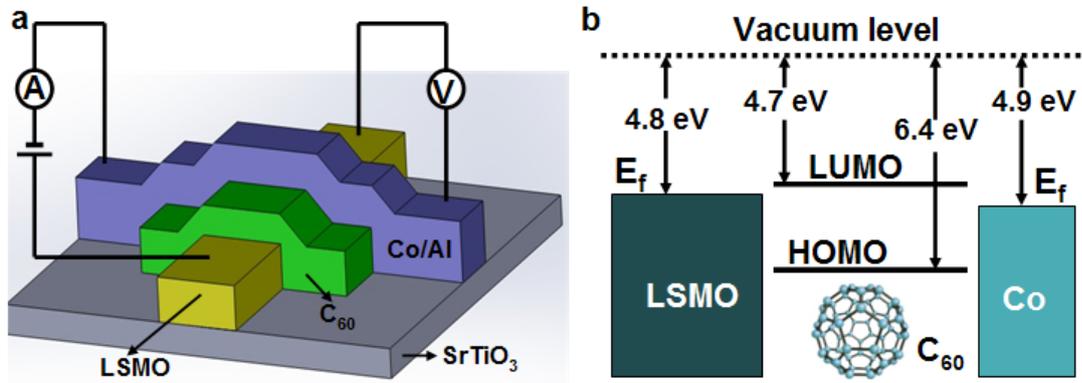

Figure 1.

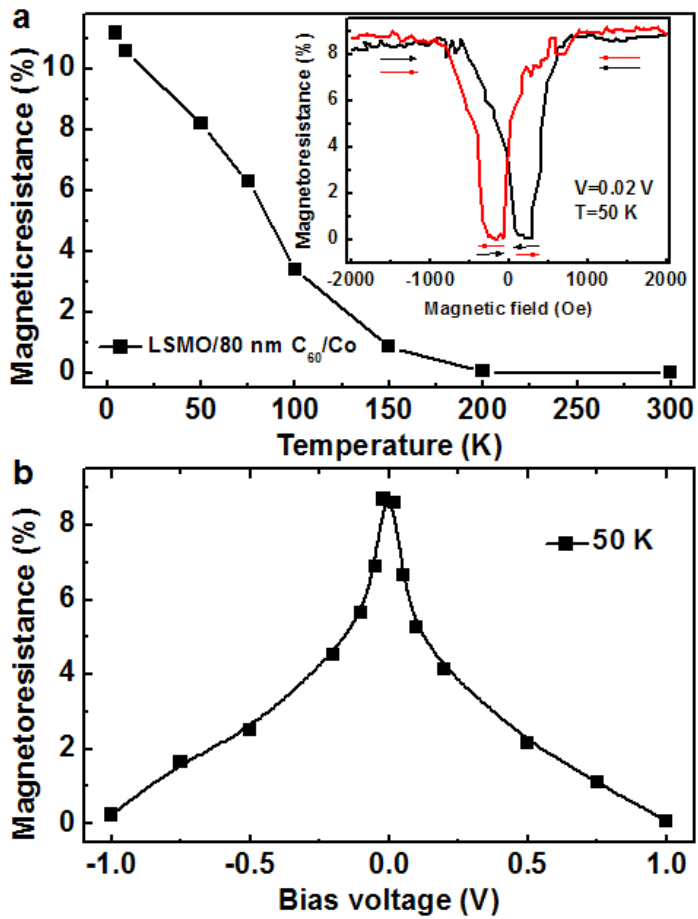

Figure 2.

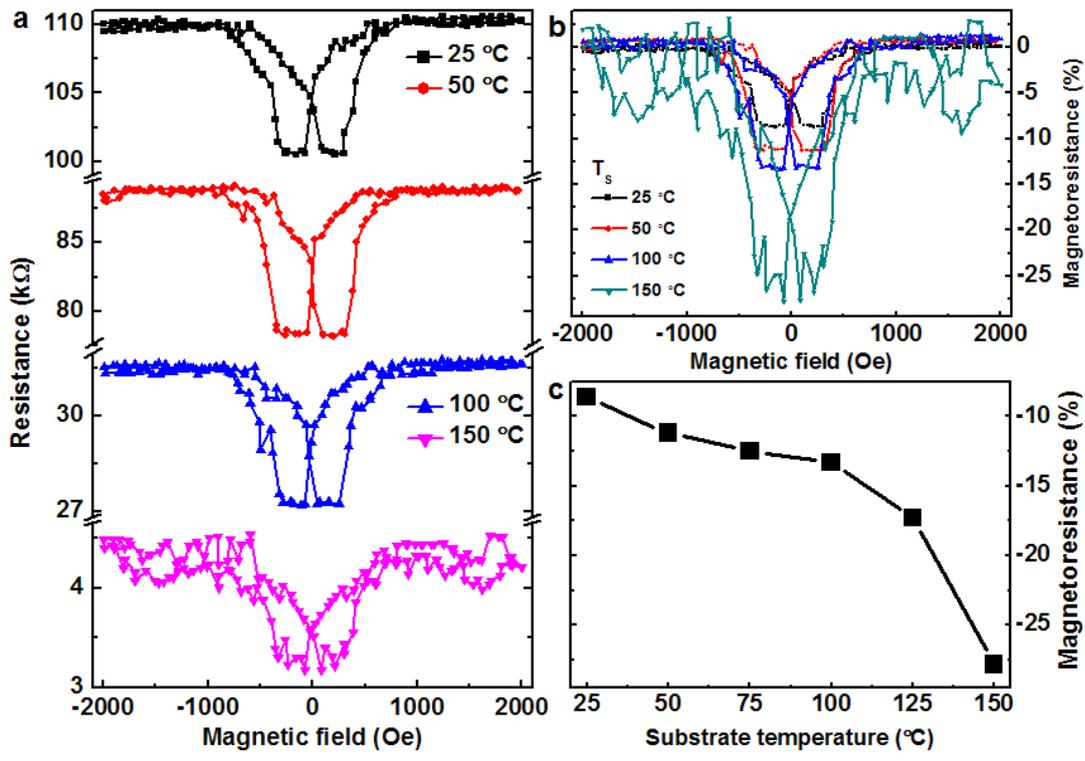

**Figure 3.**

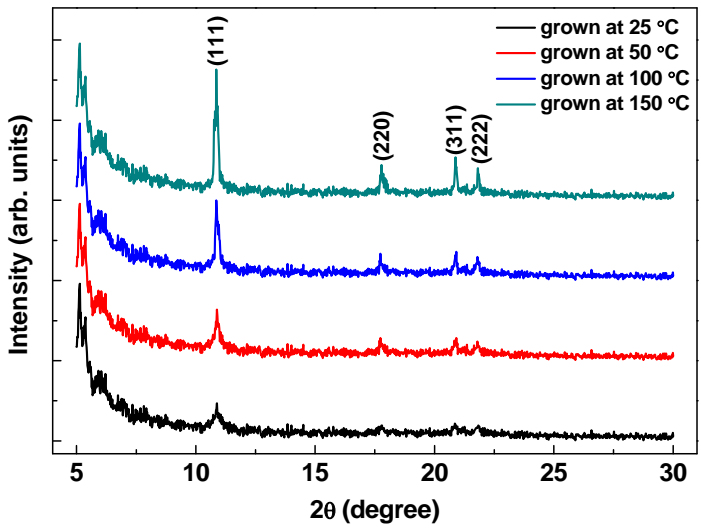

**Figure 4.**

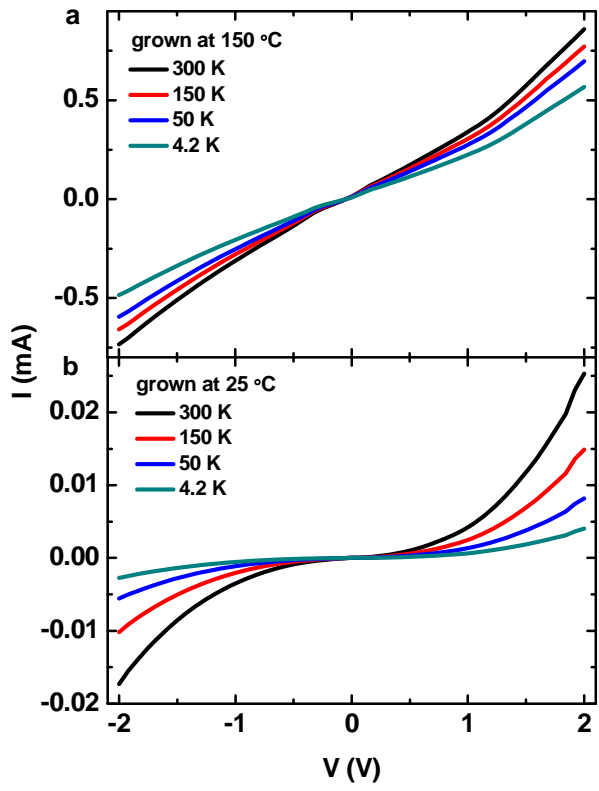

**Figure 5.**

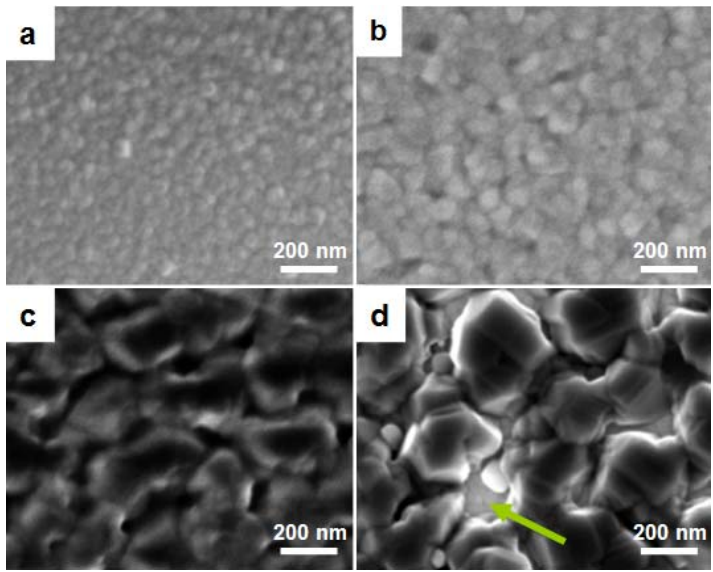

**Figure 6.**

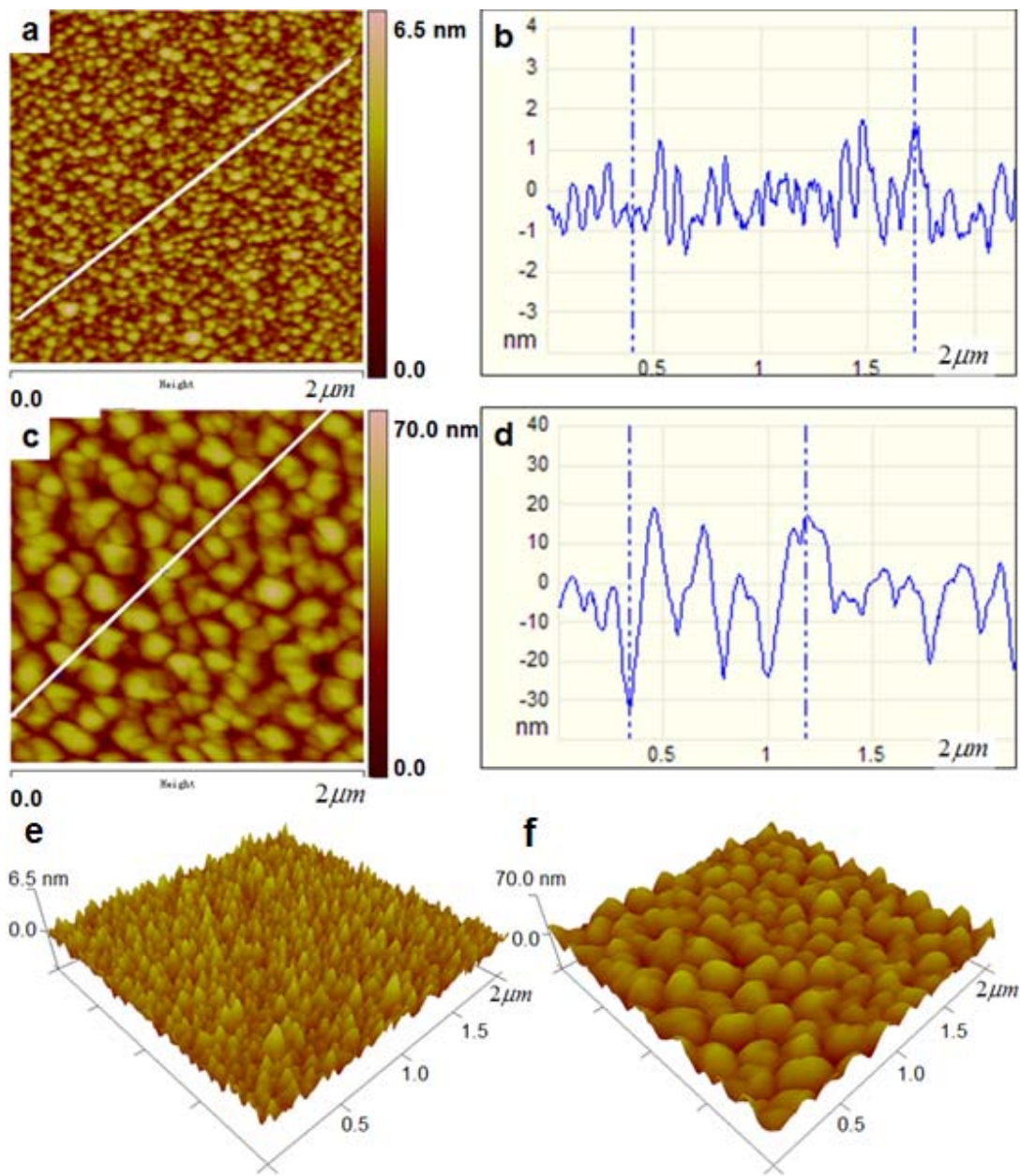

**Figure 7.**